\documentclass{tADP2e}

\usepackage{epstopdf}
\usepackage{subfigure}

\theoremstyle{plain}

\theoremstyle{definition}

\theoremstyle{remark}

\begin{document}



\title{Time-domain Studies of M31}

\author{
  \name{C.-H. Lee\textsuperscript{a}$^{\ast}$\thanks{$^\ast$Corresponding author. Email: leech@naoj.org}}
    \affil{\textsuperscript{a} Subaru Telescope, National Astronomical Observatory of Japan}
\received{v5.0 released January 2015}
}

\maketitle

\begin{abstract}
  M31, our closest neighboring galaxy, is a stepping stone to studies of stellar evolution, star formation, galaxy evolution, and cosmology. However,
  due to the difficulties of performing photometry in such crowded fields and the lack of wide cameras to encompass the entire galaxy, there has not been
  a complete census of the stellar contents of M31.
  The advent of wide-field camera
  provides us a unique opportunity to have a complete view of our neighboring galaxy, enabling an inventory of its variable content. We present a
  review of recent progresses of wide-field,
  high cadence surveys of M31, covering different population of variables and transients. We also outline future studies enabled by on-going and up-coming facilities.
\end{abstract}


\section{Introduction}
Transients and variables in the local group play pivotal roles in our understanding of stellar evolution, star
formation history, galaxy formation, and cosmology. As the nearest spiral galaxy, M31 provides us a
unique opportunity to study transients and variables in detail. The merits of M31 are: i) it has a simple
geometry and we can assume all the transients and variables are at the same distance;
ii) most of the M31 varying stars are bright enough to be resolved; moreover, we can even resolve the
progenitors of transients with the exquisite spatial resolution of the Hubble Space Telescope\footnote{Deep observations with the Hubble Space Telescope are available via the Hubble Source Catalog \cite{wab16}, which may complement time-domain studies.};
iii) in contrast to LMC and SMC which are metal poor, M31 is metal rich thus can serve as a local counterpart of the spiral galaxies that are used to determine the extra-galactic
distance \cite{fmg01}; and iv) it is a local benchmark to calibrate the Tully-Fisher relation.
Despite its proximity, there has not been a census of variables in M31
due to its extended disc
(spanning more than 2 degrees in diameter). With the advent of ultra-wide camera, e.g.
the MegaCAM on-board the Canada-France-Hawaii Telescope (CFHT), the Giga Pixel Camera on-board the Pan-STARRS1 (PS1), and the Hyper
SuprimeCAM (HSC) on-board the Subaru telescope, it is possible to have an M31 variable inventory, with periods
ranging from intra-day to several years. Further more, it is also possible to conduct studies of M31 transients,
with time scales as short as few minutes.
In this review, we aim to provide an overview of the public M31 data,
mainly gathered by Isaac Newton Telescope (INT) \cite{aeh04,vrj06}, CFHT \cite{fv12}, PS1 \cite{lrk12} and Palomar Transient Factory (PTF) \cite{lny13}. We also discuss the scientific outputs derived from these data, to demonstrate the importance of continuous monitoring
of nearby galaxies. We then discuss the prospects of M31 variability studies using future ground-based and space-based facilities, such as HSC \cite{mkn12} and Wide Field Infrared Survey Telescope (WFIRST) \cite{caa13}.
This review is organized as follows. In section \ref{sec.surveys}
we present an overview of the existing public M31 time-series data.
In sections \ref{sec.results} - \ref{sec.results3}, we outline the science cases delivered by the time-domain studies of
M31, including periodic, aperiodic variables and transients, followed by prospects in section \ref{sec.future}.

\section{Previous and on-going M31 surveys}
\label{sec.surveys}
\subsection{M31 as distance anchor}
\begin{figure}
  \begin{center}
    \includegraphics[width=\columnwidth]{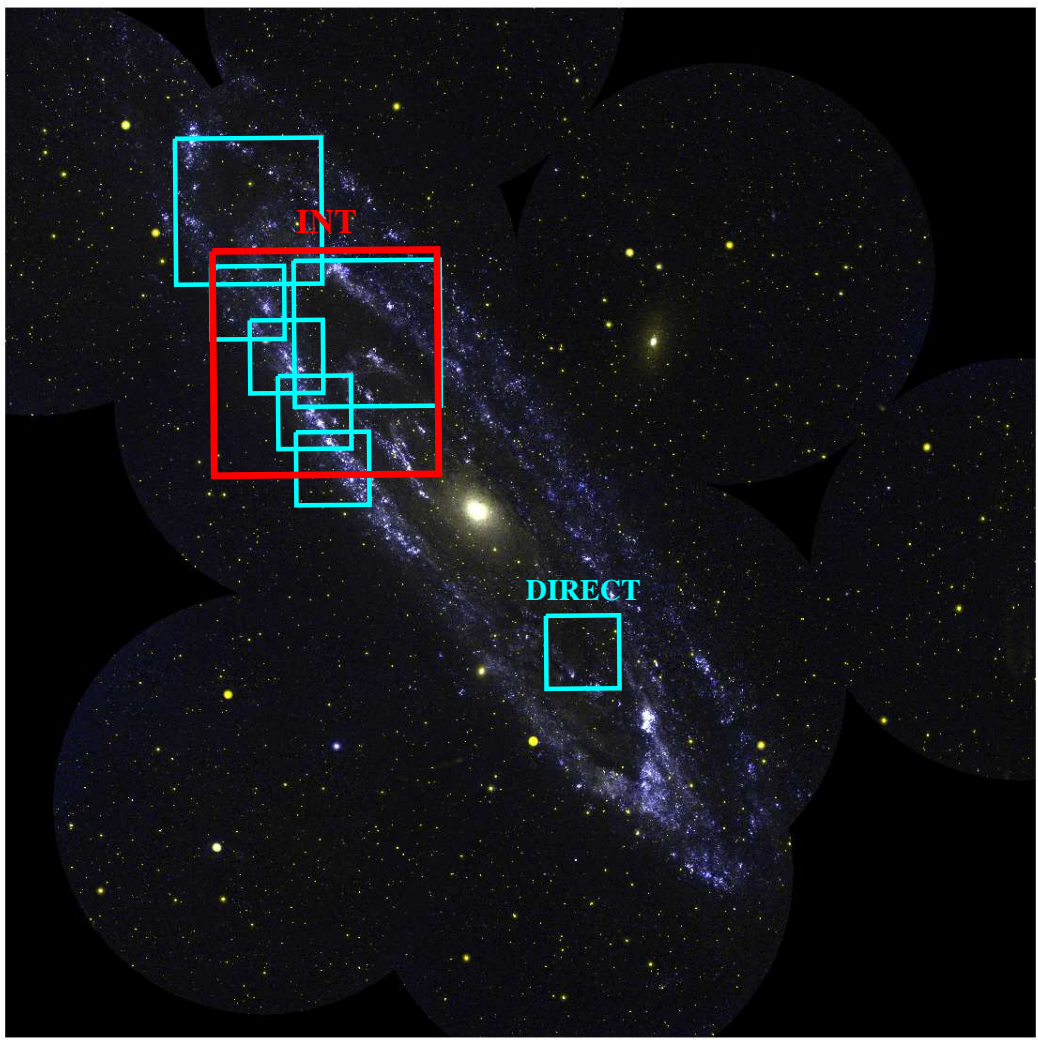}
    \caption{Footprints of distance indicator surveys -- the DIRECT (cyan squares) and the WFCAM/INT (red square) surveys. The underlying
    image comes from the UV space telescope Galaxy Evolution Explorer (GALEX \cite{gbm07}).}
  \label{fig.direct}
  \end{center}
\end{figure}

\subsubsection{DIRECT project}
As one of the only two nearby spiral galaxies in the local group, M31 is a stepping stone
to establish extra-galactic distance scale. However, the uncertainty of its distance remained at 10\% or larger in the 1990s. With the goal to provide an accurate calibration of the
extra-galactic distance scale, the DIRECT project was the first attempt to systematically search for
and characterize of eclipsing binaries and Cepheids in M31 (and M33) \cite{ksk98,skk98,skk99,kms99,mks99,mss01,mks01a,mks01b,bss03}. Using the 1.2-m telescope at the F. L. Whipple Observatory (FLWO) and the 1.3-m Michigan-Dartmouth-MIT (MDM) telescope with
a field of view (FOV) of 11$\times$11 arcmin$^2$, the DIRECT project
observed several regions in M31 disc (mainly the northern part, see Fig. \ref{fig.direct})
from September 1996 till November 1999, delivering light curves with
$\sim$ 200 epochs. This led to the discovery of 89 eclipsing binaries and 332 Cepheids
in M31 \cite{bss03}. Though the light curves from the DIRECT project suffered from severe blending effects
due to the poor site seeing, they provide a good constraint on the timing of the binary eclipses.
The variables' light curves in B, V, and I-band can be downloaded from the DIRECT project
web-site\footnote{http://www.astronomy.ohio-state.edu/~kstanek/CfA/DIRECT/}.

\subsubsection{WFCAM/INT searches for eclipsing binaries and Cepheids}
To obtain accurate distance of M31, Ribas et al. \cite{rjv04} carried out a deep photometric observations of the
northern half of M31 (see Fig. \ref{fig.direct}) for 21 nights between 1999 and 2003,
amounted to 255 and 260 photometric measurements in
V and B-band, respectively. Using the 34$\times$34 arcmin$^2$ Wide Field Camera (WFCAM) mounted
on the 2.5-m Isaac Newton Telescope, they were able to reach objects as faint as V=24 mag, with photometric
accuracy of 0.002 mag at V=18 mag and 0.2 mag at V=24 mag, respectively. This led to the discovery of 437 eclipsing binaries
and 416 Cepheids. Vilardell et al. \cite{vrj06} selected 24 bright eclipsing
binaries with photometric errors at 0.01 mag level that are promising candidates to spectroscopically follow-up with 8-m 
telescopes. Ribas et al. \cite{rjv05} and Vilardell et al. \cite{vrj10} provided follow-up of two eclipsing binaries and
determined M31 distance to 4\% level. Vilardell et al. \cite{vjr07} also presented a comprehensive studies of the Cepheid
variables. However, due to the blending effect, they derived an M31 distance modulus with rather large
uncertainty. All WFCAM/INT light curves of the variables in B- and V-band can be retrived from VizieR Online Data Catalog
associated with the work of Vilardell et al. \cite{vrj06}\footnote{http://vizier.u-strasbg.fr/viz-bin/VizieR?-source=J/A+A/459/321}.

\subsection{Searching for MACHOs}
\begin{figure}
  \begin{center}
    \includegraphics[width=\columnwidth]{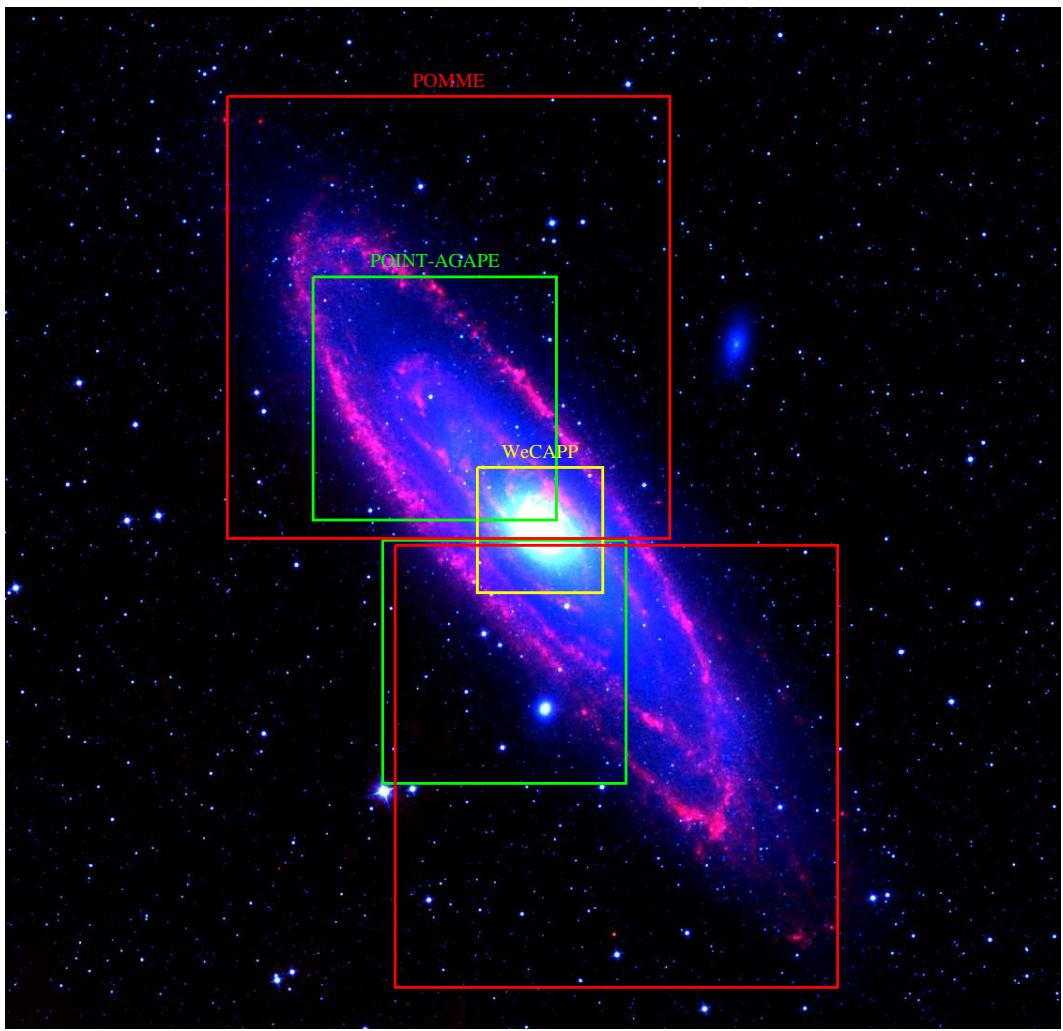}
    \caption{Footprints of high-cadence microlensing surveys -- the POINT-AGAPE (green squares), WeCAPP (yellow square),
      and POMME (red squares) survey. The underlying image
      comes from the infrared space telescope Wide-field Infrared Survey Explorer (WISE, \cite{wem10}).}
  \label{fig.ml}
  \end{center}
\end{figure}

\subsubsection{POINT-AGAPE}
The Pixel-lensing Observations with the Isaac Newton Telescope -- Andromeda Galaxy Amplified Pixels Experiment (POINT-AGAPE)
was a survey that monitored both the northern and southern discs of M31 to search for gravitational microlensing events.
Using the Wide-Field Camera mounted on the 2.5-m Isaac Newton Telescope, two 34$\times$34 arcmin$^2$ fields, in
the northern and southern part of M31's disc (see Fig. \ref{fig.ml}), were observed in Sloan g', r', and i'-bands for 60 nights
between 1999 and 2001. This data-set rendered the
discovery of a handful of microlensing events, 20 Novae and 35,414 variables. The catalog of the variables can be
obtained from VizieR Online Data Catalog associated with the work of
An et al. (2004, \cite{aeh04})\footnote{http://vizier.u-strasbg.fr/viz-bin/VizieR?-source=J/MNRAS/351/1071}.

\subsubsection{WeCAPP}
The WeCAPP project \cite{rfg01} was a dedicated survey to search for microlensing events towards the central region of M31. It continuously monitored the bulge of M31 (see Fig. \ref{fig.ml}) between September 1997 and March 2008 using the 0.8 m telescope at the Wendelstein Observatory in the Bavarian Alps. The data were taken optimally on a daily basis in both R and I filters with a field of view of 8.3 $\times$ 8.3 arcmin$^2$. From June 1999 to February 2002, WeCAPP further incorporated observations with the 1.23-m telescope of the Calar Alto Observatory in Spain, with a FOV of 17.2 $\times$ 17.2 arcmin$^2$. After 2002 WeCAPP solely used the Wendelstein telescope to mosaic the full Calar Alto field of view with four pointings. As a by-product, WeCAPP also detected 23,781 variables in the nucleus region
of M31. The catalog of these variables can be obtained from VizieR Online Data Catalog associated with the work of Fliri et al. (2006, \cite{frs06})\footnote{http://vizier.u-strasbg.fr/viz-bin/VizieR?-source=J/A+A/445/423}.

\subsubsection{POMME}
The Pixel Observations of M31 with MEgacam (POMME, PI Magnier) obtained exquisite data of M31 in 2004
and 2005 with the 1 degree$^2$ FOV MegaCAM on-board CFHT. Thanks to the large field-of-view of MegaCAM,
most of the disc of M31 can be covered with merely two pointings (see Fig. \ref{fig.ml}).
The observations were carried out in three filters
(Sloan g, r, and i-band) for up to 50 epochs in each filter. The individual exposure amounted to
400 seconds in r and 520 seconds in g and i, with stacked images
as deep as 25-26 mag in r-band and with median seeing between 0.7
and 0.8 arcsec. Using this data-set, Fliri \& Valls-Gabaud
(2012, \cite{fv12}) identified more than 2500 Cepheids, and Riess et al. (2012, \cite{rfv12}) combined the
exquisite HST infrared photometry and the periods from the POMME Cepheids to derive a clean PL
relation in the infrared. The light curves and the variable star catalog have not been released yet. Nevertheless
the reduced images can be downloaded from the CFHT Science Archive\footnote{http://www.cadc-ccda.hia-iha.nrc-cnrc.gc.ca/en/cfht/}.

\subsection{Surveys with ultra-wide cameras}
\label{sec.ultrawide}
\subsubsection{Pan-STARRS 1}
Starting from 2010, the 1.8-m Pan-STARRS 1 telescope had devoted 2\% of its total observing time to routinely monitor
the entire M31 disc when it is visible during the 2nd half of the year.
The observations were carried out mainly in r and i-band, with intra-night
cadence aiming to discover microlensing events. With the $\sim$ 7 deg$^2$ FOV, PS1 can image the entire disk of M31, as well as
the companion galaxy M32 and NGC 205 in a single shot. PS1 accumulated $\sim$ 300 epochs by the end
of 2012. The median seeing in each filter was $\sim$ 1 arcseconds. Such high cadence data
not only allowed to discover microlensing events, but also yielded high quality light curves to
study variables.  
Using this data-set, the PS1 team presented $\sim$ 2000 Cepheids \cite{krh13} and
$\sim$ 300 eclipsing binaries \cite{lks14}. The catalog of the Cepheids can be found at the VizieR Online Data Catalog, associated with the work of Kodric et al. (2013, \cite{krh13}). The full catalog and reduced images of PS1 will be released in the near future.

\subsubsection{Palomar Transient Factory}
The Palomar Transient Factory (PTF, \cite{rkl09}), currently in the second phase -- intermediate PTF -- will enter the third phase (ZTF) in 2017, is an ambitious survey to discover transients in the Universe. It utilizes the 48-inch Samuel Oschin Telescope (also known as P48) at Palomar Observatory, equipped with the CFH12K mosaic
camera redesigned to fit the telescope, with a pixel size of 1.01 arcsec/pixel and a FOV of 7.26deg$^2$. It routinely patrols the northern
sky with cadence as short as 60-seconds, with emphasis on the local luminous regions like M31.
Intriguing transients will be followed up by the dedicated P60 telescope for light curves and spectra. Though the site seeing is not
optimized to resolve the stellar populations in M31, with image subtraction method it is possible to obtain decent light curves
at least for the bright variables. M31 has been substantially observed by P48 since the beginning of 2009 with a time resolution of up to 1 day. The long-term monitoring from PTF also provides light curves with very long base line, especially
useful for studies of long-period variables \cite{nly15}. The reduced images and catalogs of PTF/iPTF can
be accessed via NASA/IPAC Infrared Science Archive\footnote{http://irsa.ipac.caltech.edu/Missions/ptf.html}.


%
%
%

\section{Scientific outputs 1: periodic variables}
\label{sec.results}
\begin{figure}
  \begin{center}
    \centering
    \includegraphics[scale=1.2]{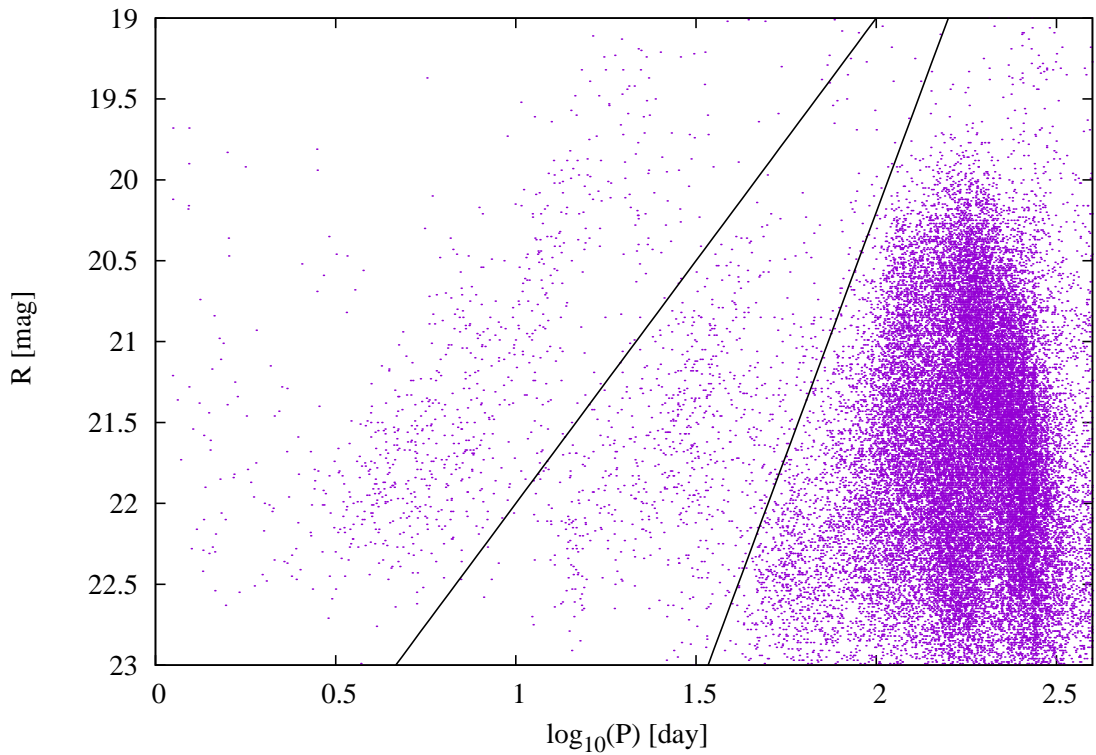}
    \caption{Variables from POINT-AGAPE surveys. The two boundaries are R[mag] = 25 - 3log$_{10}P$ and R[mag] = 32.2 - 6log$_{10}P$,
    separating the variables into classical Cepheids (left), population II Cepheids (middle) and LPVs (right) in the period-luminosity plot.}
  \label{fig.var}
  \end{center}
\end{figure}

The past and on-going M31 surveys have identified immense amount of variables. For illustrational purpose, we show the period-luminosity diagram
from POINT-AGAPE survey in Fig. \ref{fig.var}. These variables can be roughly categorized into classical Cepheids, population II Cepheids,
and long-period variables (LPV) like Mira and semi-regular variables. These variables are not only valuable to study the stellar
evolution, they can also serve as distance indicators, as well as star formation and metallicity tracers. In the following we discuss some aspects of
these periodic variables, putting emphasis on Cepheids and eclipsing binaries.

\subsection{Cepheids}

\subsubsection{Period luminosity relation}

Cepheids are pulsating variables that periodically change their radius and effective temperature. Their tight period-luminosity
(PL) relation (also known as the Leavitt law \cite{lp12}) makes them ideal distance indicators. For example, Hubble (1929, \cite{h29}) used
Cepheids to determine the distance of M31 and establish its extra-galactic nature, settling the Great Debate between Shapley and Curtis in the 1920s. Because of their tight PL relation, Cepheids have played an important role in the cosmological distance scale,
partly helping the discovery of the dark energy and the expanding universe.
Our knowledge of Cepheid PL relation mainly anchors to three galaxies, i.e. the Milky Way, the Large Magellanic Cloud, and the
maser galaxy M106. In the Milky Way currently we only have a dozen of Cepheids with accurate distances from parallaxes\footnote{With the new spatial scanning mode of Hubble Space Telescope developed by Riess et al. (2014. \cite{rca14}) and Casertano et al. (2016. \cite{cra16}), as well as the all-sky surveying space 
  telescope Gaia, the number of Cepheids with accurate parallax distance will increase dramatically in the near future.}. The
LMC distance is accurately known (to 2\% level) and it hosts numerous Cepheids, but it is metal poor compared to the SNe Ia host galaxies,
and the metallicity influence on the Cepheid PL relation is under debate. The distance to the third anchor, M106, though 
determined independently from its geometric maser, still has a 3\% error \cite{hrm13}. In addition, Efstathiou (2014, \cite{e14})
pointed out that its maser distance is in tension with the distance determined by PL relation based on the Milky Way Cepheids.
In this regard, it is important to establish other independent distance anchors. As the closest spiral galaxy, M31 is an ideal
target to become another distance anchor, mainly because its metallicity is similar to SNe Ia host galaxies and its geometry effect is negligible compared to the LMC. Its proximity also enables us to resolve various stellar populations; for Cepheids, we
can even estimate the crowding effect thanks to the exquisite spatial resolution of Hubble Space Telescope.

The current largest public M31 Cepheid sample comes from the PS1 survey, where Kodric et al. (2013, \cite{krh13}) presented the
r' and i'-band light curves of 2009 Cepheids in M31, including 1440 classical Cepheids, 126 Cepheids in the first overtone mode, 147 population II Cepheids, and 296 un-classified Cepheids. We note that although the POMME survey presented the results of more
than 2500 Cepheids, unfortunately the coordinates, magnitudes, and periods of their sample have not been public yet.

\begin{figure}
  \begin{center}
    \includegraphics[width=\columnwidth]{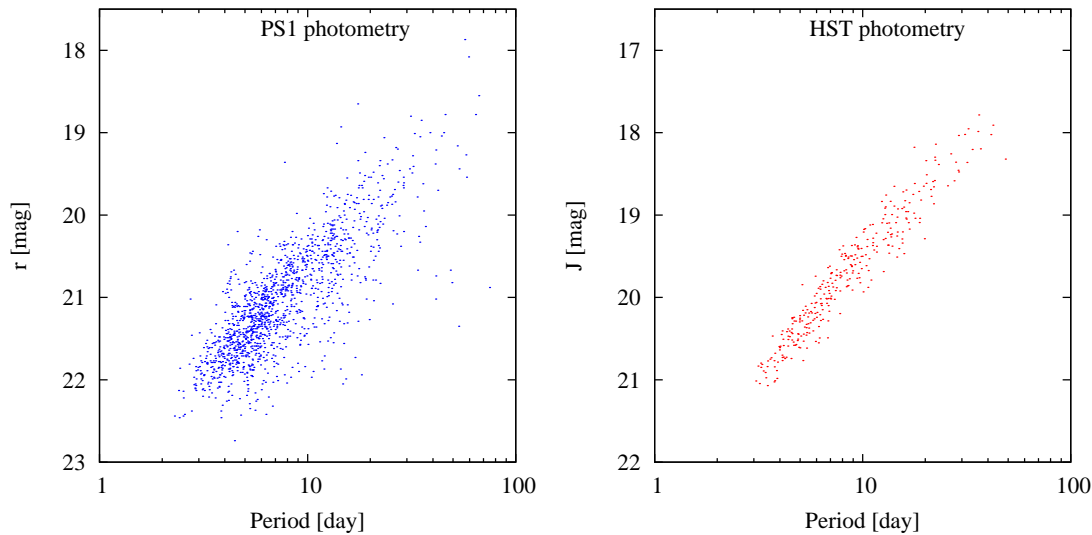}
    \caption{Classical Cepheids PL relation. Left: r-band mean-phased photometry from ground-based PS1 telescope. The photometry is heavily effected
      by the blending effects. Right: J-band random-phased photometry from Hubble Space Telescope. The exquisite spatial
      resolution delivered by HST enables us to get rid of the blending effects. The Cepheids light curve amplitude is relatively
    small in the infrared than in the optical band, so even random-phased photometry yields very clean PL relation.}
  \label{fig.hstpl}
  \end{center}
\end{figure}

Although the ground-based Cepheid photometry suffers from heavy blending
effects due to the poor seeing on image, with the add of the exquisite spatial resolution of HST from space, and taking the
advantage that the Cepheid pulsational amplitude is very small in the infrared, we can establish a clean PL relation in
the infrared, as first demonstrated by Riess et al. (2012, \cite{rfv12}). Another advantage of using infrared photometry is
in the extinction effect is much smaller (6$\times$ smaller) than in the optical bands. In Kodric et al. (2015, \cite{krs15}) they
showed that by combining the period from ground-based telescope, with photometry from space telescope, it is not only
possible to establish a clean PL relation, but also enables us to investigate the broken PL relation of Cepheids
at period around 10 days (see Fig. \ref{fig.hstpl}).

\subsubsection{Period age relation}

There is a well-established period-age relation for classical Cepheids, e.g. Bono et al. (2005, \cite{bmc05}). A simple explanation is assuming Cepheids follow the PL relation,
the mass-luminosity relation, and the stellar ages predicted by the evolutionary models, than a Cepheid with a longer period implies a higher mass,
which, from the evolutionary models' perspective, indicates a younger age. Given the period-age relation, we can have a relative age estimate of the
Cepheids. As classical Cepheids originate from young stellar population, they can be used to trace the recent star formation history. Indeed, when
investigating the spatial distribution of the classical Cepheids from PS1, they align along the 10-kpc star forming ring reported by Davidge et al. (2012 \cite{dmf12}). Taking 
a closer look, there is a age gradient along the 10 kpc ring, where the further outwards of the ring, the older the Cepheids are. This age gradient is also
in good agreement with the Spitzer observations of M31 \cite{gbe06}, supporting the suggestions that such ring-like structure is formed from
M31-M32 interaction.

\subsubsection{Ultra-long period Cepheids}

The ultra-long period Cepheids (ULPCs) are rare classical Cepheids that pulsates at period longer than 80 days. They are at the bright end of the PL relation, but appears to follow a flatter PL relation, hence are conventionally discarded when constructing PL relation. Nevertheless, because the ULPCs are intrinsically luminous (M$_{I}$ = -7.9 mag), they can provide distance estimator
up to 100 Mpc using current and future space-based facilities, providing a single-step distance measurement to galaxies in the Hubble flow.
Currently most of the ULPCs are discovered in metal-poor environments, such as dwarf galaxies or irregular galaxies \cite{bsp09}. Recent works \cite{fcm12}
provided an update of 37 ULPCs in 10 galaxies with a wide metallicity range (7.2$<$12+log(O/H)$<$9.2 dex), yet neither the dwarf and irregular galaxies, nor the spiral galaxies have accurate distance.
In this regard, M31 -- with accurate distance -- provides a unique opportunity to investigate the PL relation of ULPCs. In Ngeow et al. (2015, \cite{nly15}) they exploited the
long-term M31 monitoring images from PTF, and reported the discoveries of 2 M31 ULPCs.
\subsubsection{Beat Cepheids}
Beat Cepheids are Cepheids that simultaneously pulsate in two radial modes. Studies of beat Cepheids can trace back to Oosterhoff's work in 1950s \cite{o57a,o57b}, who introduced the second period to explain the large scatters in the light curves of U TrA and TU Cas. There were several
efforts to identify Milky Way beat Cepheids \cite{pa79,h79,h80},
however, only 20 Galactic beat
Cepheids are known to-date (see, e.g., the McMaster Cepheid Data Archive\footnote{http://crocus.physics.mcmaster.ca/Cepheid/}). The extra-galactic samples of beat Cepheids mainly came
from microlensing surveys. For example, the MACHO project reported 45 beat Cepheids in LMC \cite{aaa95}. The OGLE-II team,
on the other hand, found 93 beat Cepheids in the Small Magellanic Clouds \cite{uss99} and 76 beat Cepheids in
the Large Magellanic Clouds \cite{sus00}. A recent study from the EROS team increased the number of beat Cepheids
in the Magellanic clouds to more than 200 \cite{mbb09}. The OGLE-III survey has identified the largest number of
beat Cepheids so far: 271 objects in the LMC \cite{spu08} and 277 in the SMC \cite{spu10}. These numbers will
increase in the future with the on-going OGLE-IV survey.

\begin{figure}
  \begin{center}
    \includegraphics[width=\columnwidth]{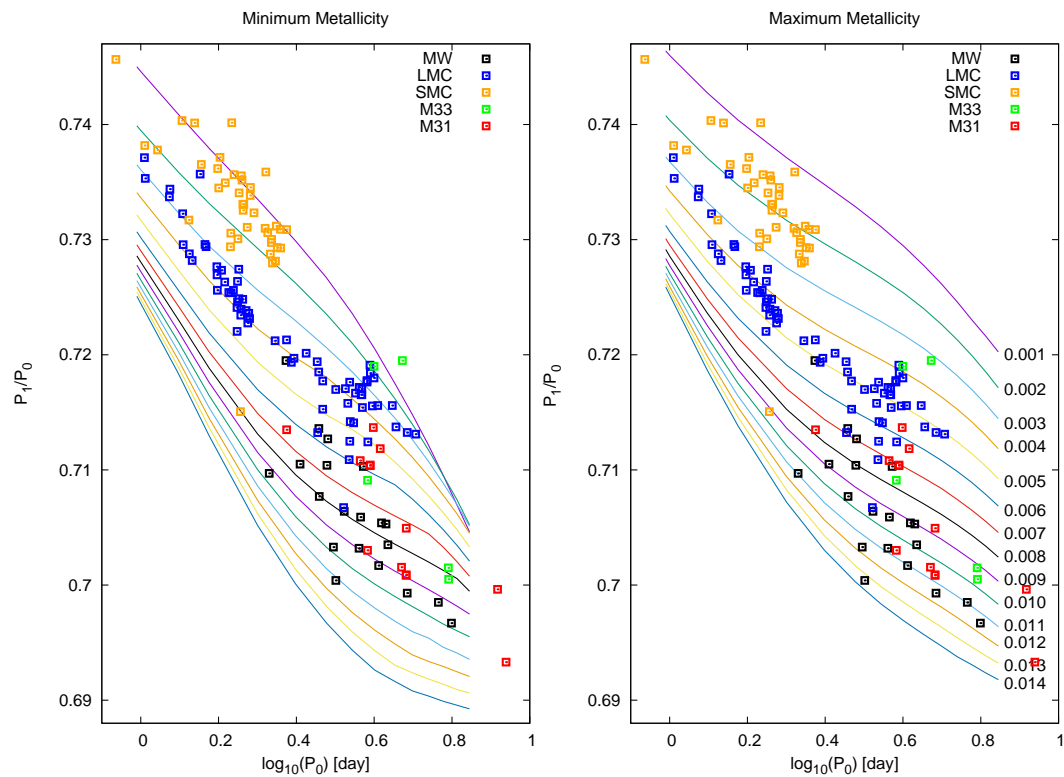}
    \caption{Beat Cepheids in M31, M33, Milky, and Magellanic Clouds. Left panel: period v.s. period ratio of the beat Cepheids over-plot
      on the lowest possible metallicity tracks. Right panel: period v.s. period ratio of the beat Cepheids over-plot on the highest
      possible metallicity tracks.}
  \label{fig.bc}
  \end{center}
\end{figure}

One of the application of beat Cepheids is they can be used to trace the metallicity content. This is because there are only four
major parameters in Cepheid pulsating model, i.e. mass, luminosity, effective temperature, and metallicity. For beat Cepheids the
two simultaneously pulsating periods provide two constraints, and we also gain an extra constraint from the mass-luminosity relation.
Since Cepheids are evolved stars crossing the instability strip, there exists only a small sub-region in the effective temperature, where
Cepheids can pulsate in two periods, thus gives us a tight constraint on its metallicity. Beaulieu et al. (2006, \cite{bbm06}) have taken advantage
of this unique application of beat Cepheids, making use of the beat Cepheids found in the CFHT M33 survey \cite{hbs06} and derived the metallicity gradient of M33 (−0.16 dex/kpc). Their metallicity gradient is in good agreement with the results derived from
HII region studies \cite{gss97}.

The disadvantage of beat Cepheids is their periods are rather short, implying that they are intrinsically very faint, hence requires
dedicated survey with decent depth to discover, especially at the distance of M31. With the intra-night cadence and deep photometry
from 1.8-m PS1 telescope, the PAndromeda team were able to identify 17 beat Cepheids in M31 \cite{lks13}. In addition, they also determine the
on-site metallicity for each beat Cepheids, and derive the metallicity gradient, which is in good agreement with results derived from planetary nebulae \cite{klb12}.

\subsection{Eclipsing binaries}

\begin{figure}
  \begin{center}
    \includegraphics[width=\columnwidth]{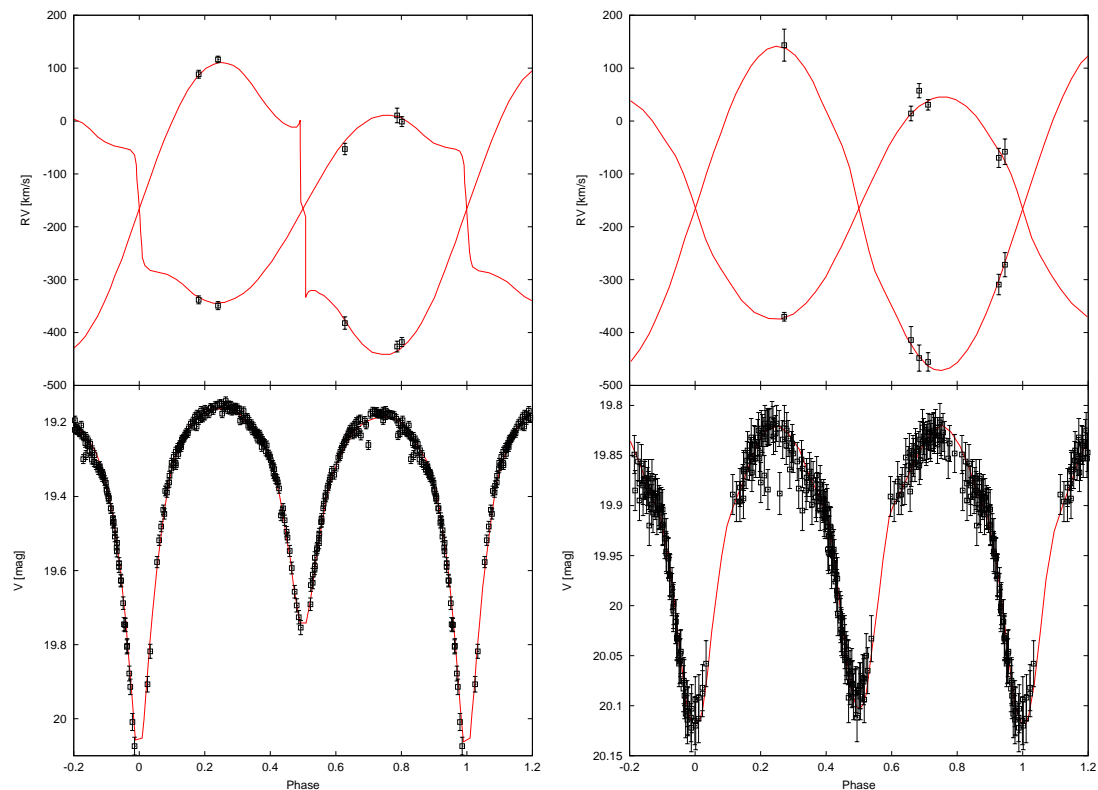}
  \caption{Radial velocity and light curves of the 2 eclipsing binaries used by Ribas et al. \cite{rjv05} (left) and Vilardell et al. \cite{vrj10} (right) to derive an independent, direct M31 distance to 4\%.}
  \label{fig.eb}
  \end{center}
\end{figure}

Paczynski (1997, \cite{p97}) has pointed out that eclipsing
binaries can serve as direct distance estimators by combining both photometric and spectroscopic information.
The photometric measurements provide us with the inclination angle, orbital period and eccentricity,
mass ratio, and radius in terms of the orbital distance. On the other hand, the spectroscopic observations
enable us to derive the mass and temperature of individual stars, as well as their orbital distance. Note the
normal degeneracy of extracting both temperature and the surface gravity (log$g$) from spectra of a single star is not
an issue for eclipsing binaries, because one can determine the mass and radius independently from the radial
velocity and light curve, hence unambiguously pin down the surface gravity of the binaries without spectra. The beauty of
eclipsing binary is that one can derive the third light ratio (or the crowding effect) from light curve analysis.
The crowding effect is often hard to measure for other distance indicators (e.g. Cepheids) especially at
extra-galactic distances. Further more, one can determine the reddening towards the eclipsing binary from
SED fitting \cite{bcm11}. The reddening is usually not measurable for other distance
anchors (e.g. Cepheids) and one has to assume a certain reddening law (e.g. R$_\mathrm{V}$ =3.1), which may not be
universal (see e.g. Bonanos et al. 2011 \cite{bcm11}, where they obtained a R$_\mathrm{V}$=5.8, significantly differs from the value
of 3.1). Once the crowding and extinction effects are quantified, we can firmly determine the distance by
comparing the inferred intrinsic luminosity to the observed flux.

It has been shown that a single eclipsing binary
system can provide a M31 distance measurement accurate to 5\% level \cite{rjv05}. The only draw back of this method is
there are limited number of eclipsing binaries that are bright enough to be spectroscopically follow-up at the distance
of M31. Even with state-of-the-art 8-10m class telescopes, it requires more than 1 hour to reach sufficient spectral S/N, not to mention
to sample the full radial velocity curve. Hence currently there are only two well-studied eclipsing binary systems. Nevertheless,
with 2 binaries Vilardell et al. (2010, \cite{vrj10}) have reached a distance estimate at 4\% level. The accurate distance also renders M31 a
distance anchor; joining with previous distance anchors, i.e. Milky Way, LMC, and M106, this helps to reduce the error budget of
Hubble constant from local measurement to below 2.5\% \cite{rmh16}.



\section{Scientific outputs 2: Aperiodic variables}
\label{sec.results2}
\subsection{Luminous blue variables}
Massive stars ($>$ 8 M$_\odot$) play an important role in galaxy evolution \cite{m13}.
Their UV radiation fuels the far-IR luminosity of the galaxy via dust heating; they depose
vast amount of mechanical energy and contribute to the chemical enrichment through stellar
wind and supernova explosion. Despite their importance, their evolution and population at
various stages are poorly constrained due to their scarcity, as well as elusive, short-lived phases,
such as the luminous blue variables (LBVs). LBVs are bright massive stars at the upper left of
the HR diagram; they experience drastic mass loss via episodic eruptions on intervals of years to
decades \cite{hd94}. They are important in two aspects: on the one hand,
they cast away a large portion of mass during eruptions, thus are considered as a transitional
stage when O stars evolve towards the H depleted Wolf-Rayet stars \cite{mem11}. On the
other hand, they are suspected immediate precursor of SNe II, such as SN 1987A \cite{s07}, SN 2005gl \cite{gl09}, and 2009ip \cite{msf13}, which has
not been predicted by theoretical studies. The duration of the LBV phase and its properties
remain poorly understood due to their rarity (only 12 known LBVs in the Milky Way \cite{cla05}. Though new IR Galactic plane surveys have revealed several LBV candidates \cite{gkf10}, follow them up individually is observationally expensive. One
alternative is to identify and characterize them in local group galaxies. The advantages are: i)
compared to the Galactic LBVs, the distance to local group galaxies are better known, which can
be used to infer their luminosity; ii) multi-object, wide-field fiber spectrograph enables efficient
follow-up.

As the LBVs show variabilities in time-scales of decades to centuries, it is difficult to identify them without
long-term monitoring. Recently there are approaches to search for LBVs using other specific properties. For
example, Massey et al. (2007, \cite{mmo07}) made use the strong H$\alpha$ emissions from LBVs, and selected LBV candidates
using H$\alpha$ imaging of the entire disc of M31. Their follow-up spectra indicate that these candidates bear
spectroscopic resemblances with {\it bona fide} LBVs, and the only missing evidence to establish the LBV nature
for these LBV candidates is the photometric variability from the light curve. On the other hand, making use of the PS1 M31 data, Lee et al. (2014, \cite{lsk14})
also suggested that it is possible to identify LBV candidates from short-term variations. With the available M31 public time-series data dated back to the
microlensing studies in 1990s, it is now
possible to verify the variability of LBV candidates with baselines longer than
a decade.

\subsection{R Coronae Borealis}

\begin{figure}
  \begin{center}
    \includegraphics[width=\columnwidth]{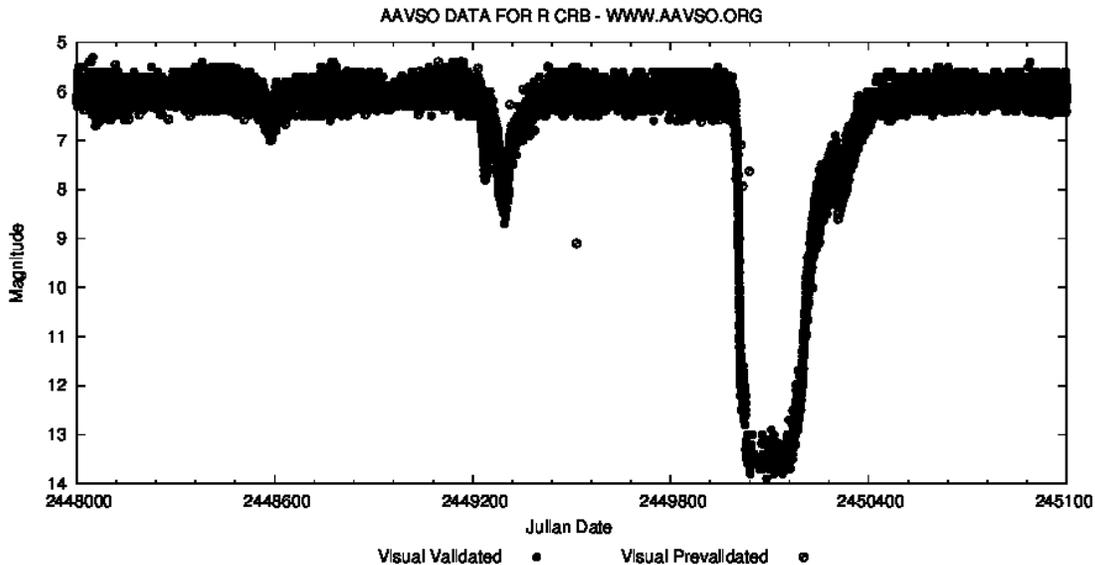}
  \caption{R Coronae Borealis light curve from AAVSO.}
  \label{fig.rcb}
  \end{center}
\end{figure}

R Coronae Borealis (RCB) stars are hydrogen deficient, carbon enriched super-giants with aperiodic photometric dimming up to 9 mag in optical bands (see Fig. \ref{fig.rcb}). The photometric variabilities are caused by sudden formations of carbon dust obscuring the photo-sphere of the star. Despite the fact that RCBs are intrinsically very bright (-5 $\le$ M$_\mathrm{V}$ $\le$ -3.5, \cite{twm09}), only 76 RCBs \cite{tww11}  have been reported in the Milky Way.
This can be attributed to the lack of all-sky, long-term survey programs to monitor their sudden dimming and/or the fact that RCBs represent an elusive phase of the stellar evolution.

The deficiency of hydrogen and enrichment of carbon in the atmosphere of RCBs suggest that they are at a final stage of stellar evolution. Currently there are two competing scenarios to explain the origins of RCBs. The first one is the white dwarf merger scenario  \cite{w84}, where RCBs are regarded as products of a carbon-oxygen white dwarf merged with a helium white dwarf, commonly dubbed the double degenerate scenario (DD). The second one is the final helium shell flash scenario (FF, \cite{ity96}), which regards RCBs as stars expanding rapidly to super-giant size shortly before turning into a white dwarf. There are several lines of evidence support each of the scenario. On the one hand, for example, observations of $^{18}$O overabundance in cool RCBs \cite{cgh07}, as well as surface abundance anomalies, especially fluorine \cite{plk08}, strongly support the DD scenario. On the other hand, discoveries of four RCBs associated with nebulae \cite{css11} indicate that RCBs can originate from the FF scenario as well. The DD scenario is of special interest because it is a low-mass analog of SNe Ia formed via the DD channel. Hence, studies of RCBs might shed lights on the progenitors to SNe Ia.

Because of our location in the Milky Way, it is difficult to conduct a complete survey for Galactic RCBs. On the other hand, M31 provides
an excellent test-beds to search for RCBs. This is because at the distance of M31, RCBs are still visible. During the sudden drop of
brightness, we may not be able to detect it at the distance of M31, nevertheless, from the disappearance and/or reappearance, we can
identify RCB candidates. In fact, Tang et al. (2013, \cite{tcb13}) have used the PTF M31 data to search for RCBs, and confirmed two RCBs, as well
as two RCB candidates, reside in M31. With the long time-baseline enabled by combining all the public archived M31 data, it is possible
to construct light curves spanning more than a decade for the majority of M31 disc, and facilitate the search for RCBs, with the
goal to conduct a statistic study of RCBs in M31.

\begin{figure}
  \begin{center}
    \includegraphics[width=\columnwidth]{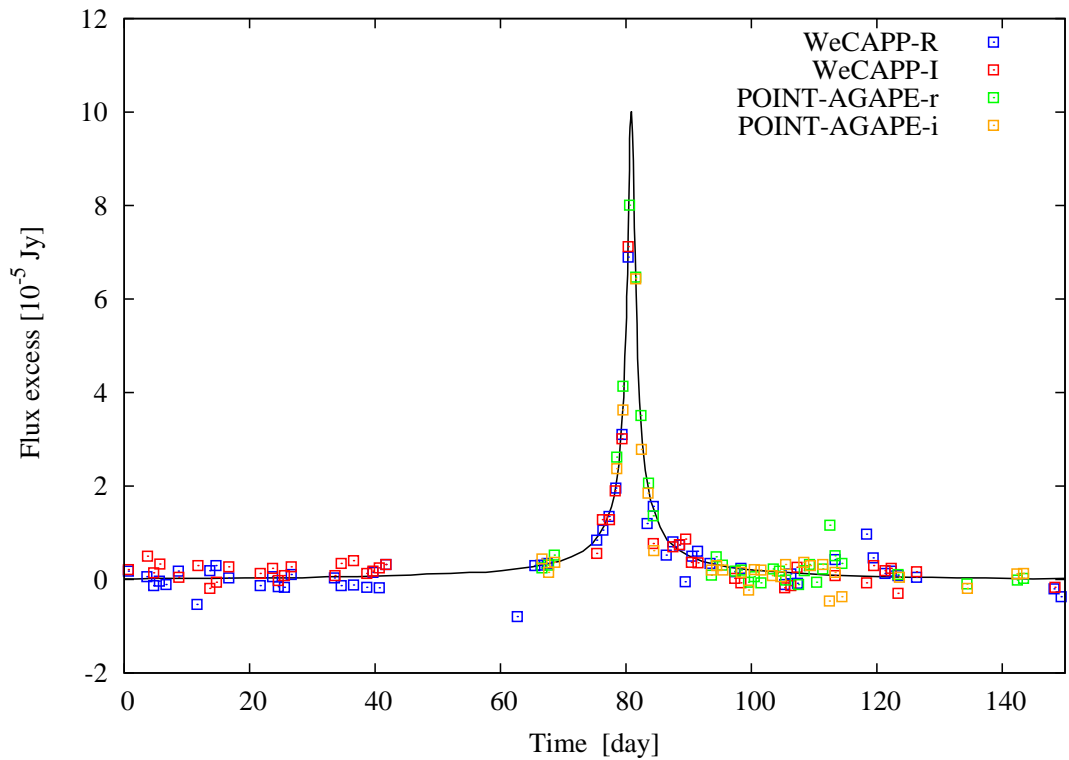}
    \caption{Microlensing light curve of POINT-AGAPE-S3/WeCAPP-1. Data points are POINT-AGAPE r-band (green), POINT-AGAPE i-band (orange),
    WeCAPP R-band (blue), and WeCAPP I-band (red). The best-fit microlensing model is over-plotted in black line.}
  \label{fig.gl1}
  \end{center}
\end{figure}

\section{Scientific outputs 3: Transients}
\label{sec.results3}
\subsection{Microlensing}
\subsubsection{Searching for MACHOs}
The existence of dark matter was first proposed by Zwicky in 1930s \cite{z33} when he used the virial theorem to calculate the
gravitational mass of the Coma galaxy cluster and recognized there are large fraction of unseen matter, which he coined
the term ``Dunkle Materie (dark matter).'' Eight decades after the discovery, the nature of dark matter is still unknown.
Dark matter can be smoothly distributed, i.e. weakly interacting massive particles (WIMPs), or in the form of massive
compact halo objects (MACHOs). As dark matter hardly produces electro-magnetic emissions, the best way to study its
properties is via gravitational interaction. Paczynski (1986, \cite{p86}) was the first one to advocate the idea of using
gravitational microlensing to detect MACHOs. Using the Magellanic Clouds as an example, Paczynski calculated the
optical depth of microlensing, i.e. at any time the probability of a star in the Magellanic Clouds is well aligned
with the lens in the Milky halo and being amplified by more than a factor of 1.34, is of order of 10$^{-6}$. This suggests
that if one monitors enough stars with sufficient cadence and time-baseline, it is possible to detect microlensing
events. This idea triggers several campaigns to search for microlensing in dense stellar
fields, e.g. the MACHO, EROS, OGLE, and MOA teams, leading to the discoveries of the first microlensing events in
the Magellanic Clouds \cite{aaa93,abb93,usk93}. During the 5.7 years of survey,
MACHO team reported 10 microlensing events, concluding that 16\% of the Milky Way halo is made up of MACHOs with masses
between 0.1 and 1 solar mass. On the other hand, OGLE-II and OGLE-III results indicate that self-lensing alone can fully
explain the microlensing events in Magellanic Clouds, and there is no need of MACHOs in the Milky Way halo.
The contradicting results partly originate from the draw back that by looking at the Magellanic Clouds, we only have
1 fixed sight-line of the entire Milky Way halo. In addition, the unknown self-lensing rate also makes the interpretation
of the microlensing results difficult.

\begin{figure}
  \begin{center}
    \includegraphics[width=\columnwidth]{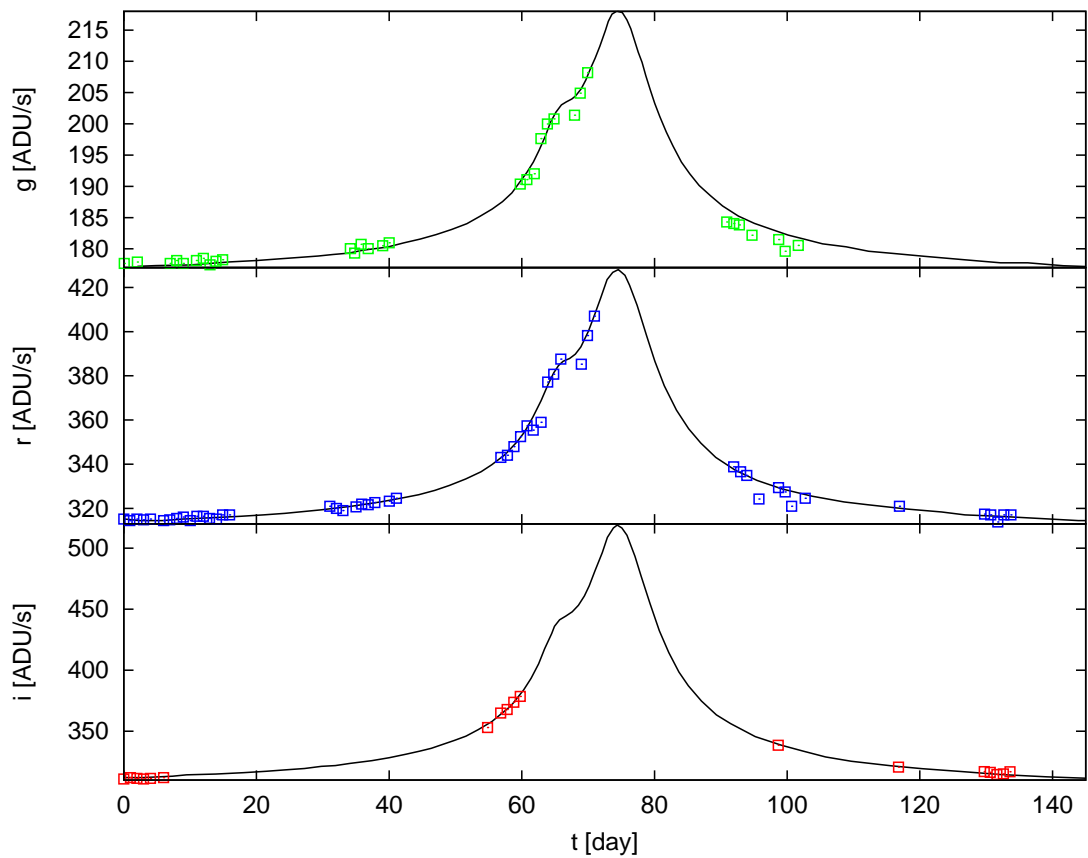}
    \caption{An M31 exoplanet microlensing candidate light curve in g' (upper), r' (middle), and i'-band (bottom panel), over-plotted with the
      best-fit binary lens model C1 with a mass ratio of 0.014 \cite{aek04}.}
  \label{fig.p99n2}
  \end{center}
\end{figure}

Besides Magellanic Clouds, M31 provides another ideal dense stellar field for microlensing studies. The benefits of M31
are: i) we can sample different sight-lines toward M31, in contrast to the fixed sight-line of Magellanic Clouds; ii)
we can probe the halo of M31 as well. As the structure of M31 is well known, we can expect an asymmetry of the
microlensing event rate among the near and far side of the disc of M31. The idea of searching for microlensing in
M31 was first proposed by Crotts (1992, \cite{c92}), which triggered several long-term monitoring campaigns, e.g. POINT-AGAPE, WeCAPP,
POMME, and PAndromeda. While there are several M31 microlensing events reported, some of them are suspected to be
variables and/or self-lensing, and the fraction of MACHOs in the M31 halo is still under debate \cite{lrs15}.
Nevertheless, there are two well-studied individual events, i.e. POINT-AGAPE-S3 / WeCAPP-1 \cite{rsb08}
and OAB-N2 \cite{cdg10}, which shows that microlensing signals are hard to reconcile with self-lensing
alone, and there's need of MACHO in the M31/Milky Way halo.

\subsubsection{Exoplanets}

Another intriguing application of microlensing is to search for exoplanets. For lenses composed of a stellar mass object associated with a planetary
mass object, we can see perturbations to the symmetric single lens light curves. As this small perturbation is only sensitive to the
mass ratio and the geometry of the deflectors, it can probe exoplanets beyond the snow line, much farther than the radial velocity or transit
method. Another advantage of microlensing is it is not limited by the brightness of the host star, and can detect
exoplanets even as distant as in M31. There is an microlensing event, POINT-AGAPE N2 (see Fig. \ref{fig.p99n2}), showing significant perturbations
to the symmetric single lens light curves. With a best-fit mass ratio of 0.014, it can be reconciled with a 6
M$_J$ exoplanet \cite{icd09}). However, as it can be explained by binary stars or parallax effects as well, its exoplanet
nature is still under debate.

\subsection{Novae}
Classical novae are subclasses of cataclysmic variables which are
composed of a white dwarf interacting with a late-type
companion star. The companion continuously loses its mass through Roche
lobe overflow, forming an accretion disc around the white dwarf.
At a certain point, the continuous mass transfer from the companion induces thermo-nuclear runaway
(TNR) onto the surface of the white dwarf, leading to the nova
eruption.

\begin{figure}
  \centering
    \includegraphics[width=\columnwidth]{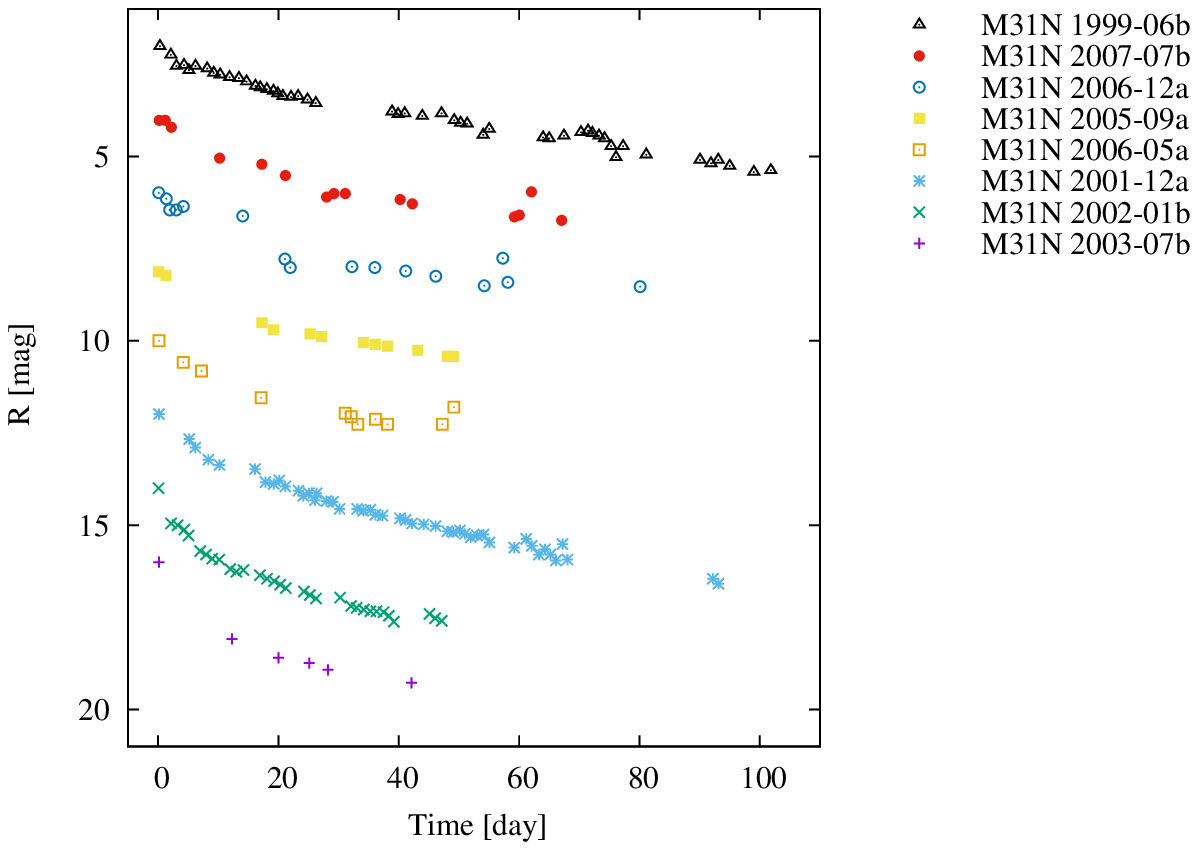}
    \includegraphics[width=\columnwidth]{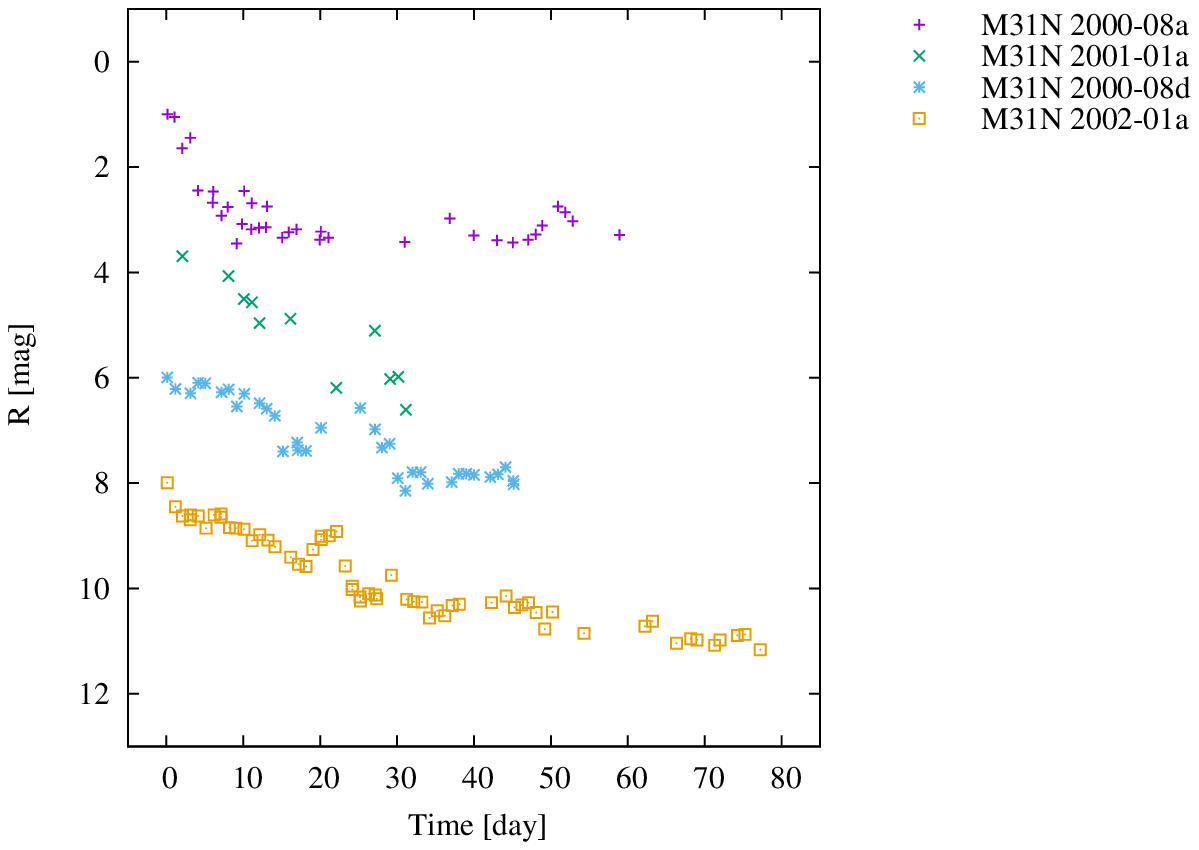}
  \caption{Upper panel: light curves of S-type novae. Lower panel: light curves of C-type novae.}
  \label{fig.snova}
\end{figure}

Novae are important stellar population tracers, potential distance indicators, and possible progenitors of supernovae.
First of all, base on the assumptions that the rate of decline is in proportional to the mass of the central white dwarf,
della Valle \& Livio (1995, \cite{dl95}) use $t_2$ -- time required to fade by 2 magnitudes below the maximum brightness -- to categorize novae into fast and slow decline classes,
and suggested that fast decline novae ($t_2<$ 12 days) are related to stars belonging to Population I with relatively massive white dwarfs, while slow novae are associated
to Population II stars and have less massive white dwarfs.

Secondly, novae have the potential to serve as standard candles of estimating extra-galactic distance. This is because there is a relation between the maximum luminosity of the light curve and the rate of decline. Hubble (1929, \cite{h29}) is the first one to notice that brighter novae have steeper decline. The empirical “Maximum Magnitude versus Rate of Decline” (MMRD) relation was further investigated by Zwicky (1936, \cite{z36}) and further studied by Mclaughlin (1945, \cite{m45}) and Arp (1956, \cite{a56}). The theoretical foundation for MMRD relation can be found in Shara (1981, \cite{s81}), with further revision by Livio (1992, \cite{l92}).

Recurrent novae are also regarded as candidate supernovae progenitors \cite{s10}. The idea is after several nova explosions, recurrent novae accumulate enough mass onto the central white dwarf envelope, and eventually turn into supernova explosions. However, as recurrent novae at extra-galactic distances are too faint to be identified, and
they have eruptive time intervals in the order of decades, we still lack direct observations of recurrent novae turn into supernovae.

Due to our fixed position in the Milky Way, it is difficult to chart all Galactic novae without dedicated all-sky surveys. In addition, due to the massive dust produced during
nova eruption and the unknown extinction, it is hard to accurately determine the distance of Galactic novae. In this regard, M31 provides us an ideal test-bed to study novae, because novae in
M31 is bright enough to follow-up even with 1-m class telescopes. We can also probe novae in different galactic components, i.e. in the bulge, in the spiral arms, and even in the halo of M31,
providing a statistical study of the spatial distribution of novae. Last but not the least, M31 is close enough that we can even resolve the progenitors of the novae with the exquisite
spatial resolution delivered by HST.

There are already several M31 novae catalogs as spin-offs from high cadence M31 microlensing surveys. For example, Darnley et al. (2004, \cite{dbk04}) present 20 novae in M31 disc from the
POINT-AGAPE survey. Lee et al. (2012, \cite{lrs12}) also present novae population in the bulge of M31 using the WeCAPP microlensing survey.
Mid-infrared survey of M31 novae, with the goal to characterize the dust formation process, has also been carried out with Spitzer\cite{sbd11}.
The most comprehensive M31 novae catalog can be found
from MPE M31 Novae group\footnote{http://www.mpe.mpg.de/~m31novae/opt/m31/index.php} \cite{phs07}, which includes the position, time of eruption, speed class, spectroscopic class, and ancillary light curve and spectra if available.

\begin{figure}
\centering
  \includegraphics[width=\columnwidth]{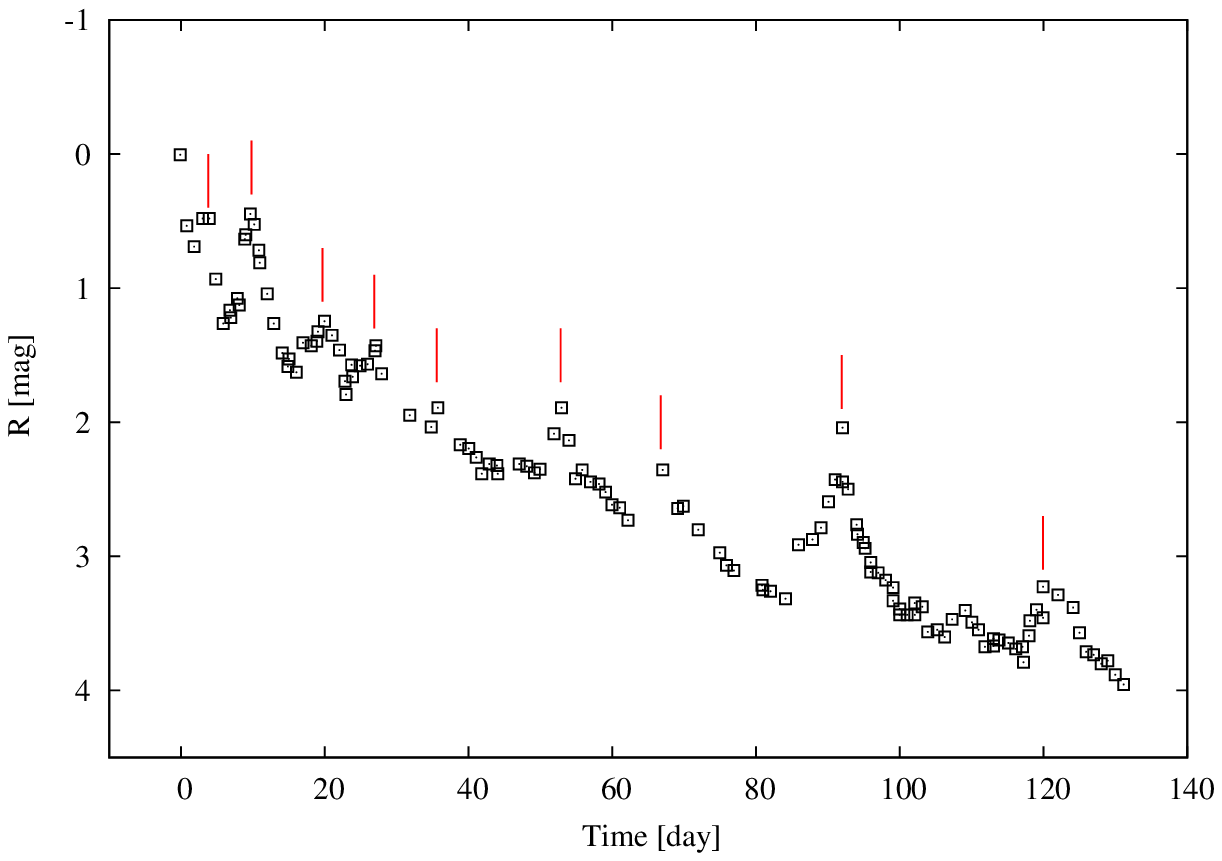}
    \includegraphics[width=\columnwidth]{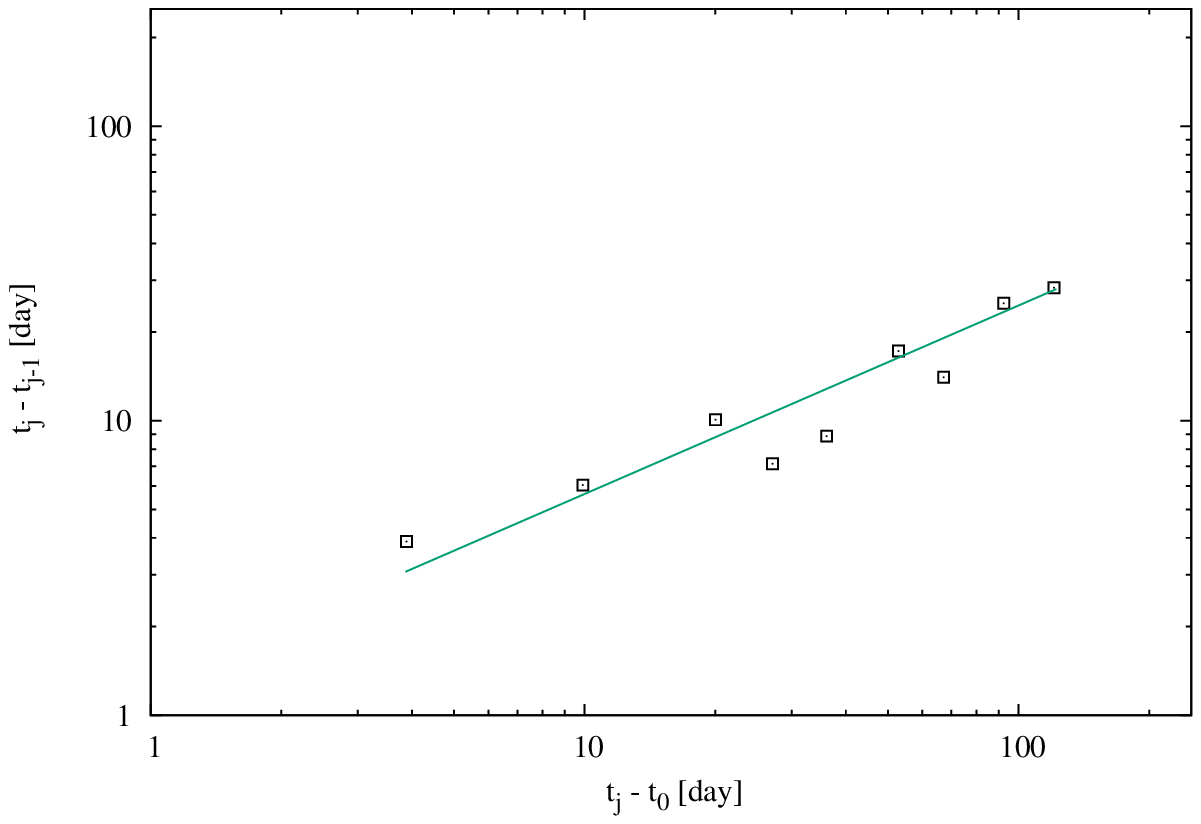}
  \caption{Upper panel: light curve of an M31 J-type nova (M31N 2001-10a). Lower panel: time intervals between the peak of successive jitters v.s. time intervals between each jitter and the time of eruption. There is a trend in the log-log plot, consistent with the hydrogen burning envelope instabilities scenario \cite{p09}.}
  \label{fig.jnova}
\end{figure}

Using the WeCAPP survey, Lee et al. (2012, \cite{lrs12}) make 91 novae light curve public. Unlike previous studies which only sort the novae according to their rate of decline, Lee et al. (2012, \cite{lrs12}) also
categorize the light curves according to the light curve morphology proposed by Strope et al. (2010, \cite{ssh10}). The public nova light curves include
many S-type (smooth) novae (see Fig. \ref{fig.snova}), enabling to build an extra-galactic novae light curve
template \cite{sg15}, as well as detailed studies of sub-type novae, e.g. novae show strong cusp feature due to immense dust formation
(see Fig. \ref{fig.snova}) and novae with jitters/flares (see Fig. \ref{fig.jnova}) which can be attributed to instabilities in their hydrogen burning envelopes \cite{p09}.

With the recently completed pan-chromatic survey of a quadrant of M31 disc by Hubble Space Telescope (PHAT, \cite{dwl12}), it is now possible to study the progenitors of
novae in M31. A pilot study of M31 novae progenitors has been conducted by Williams et al. (2014, \cite{wdb14}), where there is an excess of novae containing red giant secondaries when compared with Milky Way novae.


\section{Prospects: going deeper, shorter, and redder}
\label{sec.future}

Although the Pan-STARRS and CFHT data allow us to identify Cepheids with periods as
short as one day, they are not deep enough to reveal the faintest periodic variables, i.e. RR Lyrae, which are 25-26 mag in g-band.

\begin{figure}
  \begin{center}
    \includegraphics[width=\columnwidth]{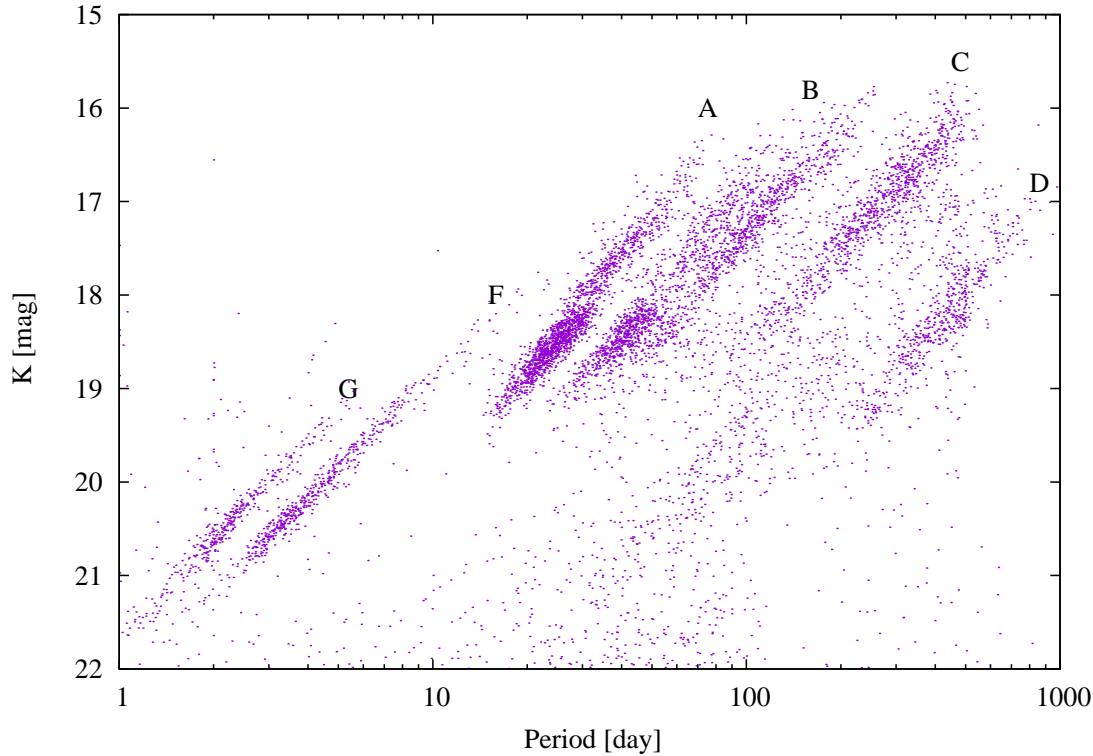}
    \caption{The infrared PL relation as seen from M31. The sample is obtained from LMC long-period variables, with
      magnitudes scaled to the distance of M31. We can see several long-period variable tracks, such as A, B, C, and D
      as shown in Ita et al. (2004, \cite{itm04}), are all brighter than K=19 mag, hence feasible with 4-m class telescopes on the
    ground. The G and F tracks mark the classical and population II Cepheids.}
  \label{fig.irpl}
  \end{center}
\end{figure}

With an age older than 10 Gry, RR Lyrae are ideal tracers of tidal trails and useful in studying
galaxy formation in the earliest stage. The periods and amplitudes of the sub-group RRab stars
are related to their metallicities \cite{sig10}, and can thus serve as metallicity
indicators of the sub-structure. Once the metallicity is determined, the absolute magnitude of the
RR Lyrae (hence the distance) can be derived accordingly \cite{cc03}.
The only uncertainty in the distance determination is the line-of-sight reddening, which can also
be inferred from the minimum light colors of the RR ab stars \cite{kcl10}. Due to
these features, RR Lyrae provide us the potential to study the tidal features and metallicities
across the disk of M31. However, due to the faintness of RR Lyrae, previous searches were
limited to sparse patches towards M31 using HST \cite{sml09,jsb11} or limited to longer period RR Lyrae with hour-long exposures from 4m class
telescopes, such as CFHT \cite{pv87} or WIYN \cite{dso04}.
Note that for RR Lyrae with shorter period, e.g. 0.2 day, one needs an hour of integration with a
4-m class telescope to detect RR Lyrae. The exposure spans 20\% of the pulsation cycle, smearing
the light curve and hampering the determination of period and amplitude. To avoid smearing of
the light curve, one needs 8-10 m telescopes to reach the required photometric depth in a
reasonable amount of time.

Recently, there is a unique wide field camera (HSC) on-board the 8-m Subaru Telescope, ideal
for the search of RR Lyrae in M31. With a FOV of 1.5 degrees in diameter, HSC can cover most
of the M31 discs with merely two pointings, providing an efficient deep survey of M31 disc. In addition,
such deep survey will also be valuable in searching for planetary microlensing events, where
the planetary perturbation of single lens microlensing light curve is very short (in the order of
intra-days).
Recently, Niikura et al. (2017, \cite{nty17}) conducted a dense cadence HSC observation of
M31 in order to search for transient stars; every exposure is taken every 2-min over about 7 hours
observations under the good seeing condition (better than 0.6'' seeing), yielding about 190 time-consecutive
images of stars over the entire M31 region. Various transient stars such as star flare, binary and variable stars are identified from the data.
The drawback of HSC is that it takes $\sim$ 30 minutes to change filters, hence makes
multi-color observations with intra-night cadence very difficult. In addition to Subaru Telescope,
one ideal telescope for M31 microlensing search is the Large Binocular Telescope; with the
two 8-m mirrors, LBT permits simultaneous multi-color observations, hence is very useful in
obtaining multi-color light curves to establish the lensing nature of microlensing events.

Another direction to advance our understanding of M31 variables is pursuing observations in the
infrared. One important factor to consider when studying variables is the foreground/intrinsic
extinction, which is negligible in the infrared wavelength compared to optical pass-bands, as
shown in the Cepheid PL relation. In addition, the long-period variables also exhibit tight
PL relation in the infrared (see Fig. \ref{fig.irpl}). As the long-period variables out-numbered
Cepheids ($\sim$20$\times$ more numerous) and are of comparable brightness in the infrared, we can
easily establish distance estimate of extra-galactic systems using a much smaller FOV. This
is important in the era of JWST, GMT, E-ELT and TMT, where most of the observations will
be carried out in the infrared, especially for the distant galaxies. In this regard, M31 can
provide a pivotal anchor for the infrared period luminosity relation. Currently there are only
static infrared images of M31, e.g. from PHAT \cite{dwl12} and from ANDROIDS \cite{scc14}. With the wide-field infrared
imager, such as WIRCam mounted on CFHT, it is possible to obtain time-series photometry of M31
disc down to K=19 mag, covering the faint end of the long-period variables. With future space-based
, wide-field facilities, such as WFIRST, it is even possible to observe the entire disc of M31
with merely 4 pointings.


\end{document}